\documentclass{article}

\usepackage[preprint]{neurips_2024}

\usepackage[utf8]{inputenc} 
\usepackage[T1]{fontenc}    
\usepackage{hyperref}       
\usepackage{url}            
\usepackage{booktabs}       
\usepackage{amsfonts}       
\usepackage{amsmath}        
\usepackage{nicefrac}       
\usepackage{microtype}      
\usepackage{xcolor}         
\usepackage{graphicx}       
\usepackage{subcaption}     
\usepackage{multirow}       
\usepackage[most]{tcolorbox} 
\usepackage{tikz}           
\usetikzlibrary{positioning, arrows.meta, fit, backgrounds, shapes.geometric, calc}

\title{CTI-REALM : Benchmark to Evaluate Agent Performance on Security Detection Rule Generation Capabilities}

\author{%
  Arjun Chakraborty \\
  Microsoft Security AI \\
  \texttt{arjunc@microsoft.com} \\
  \And
  Sandra Ho \\
  Microsoft Security AI \\
  \texttt{sandraho@microsoft.com} \\
  \And
  Adam Cook \\
  Microsoft Security AI \\
  \texttt{cookadam@microsoft.com} \\
  \And
  Manuel Mel\'endez \\
  Microsoft Security AI \\
  \texttt{mmelndezlujn@microsoft.com} \\
}

\begin{document}

\maketitle

\begin{abstract}
 CTI-REALM (Cyber Threat Real World Evaluation and LLM Benchmarking) is a benchmark designed to evaluate AI agents' ability to interpret cyber threat intelligence (CTI) and develop detection rules. The benchmark provides a realistic environment that replicates the security analyst workflow. This enables agents to examine CTI reports, execute queries, understand schema structures, and construct detection rules. Evaluation involves emulated attacks of varying complexity across Linux systems, cloud platforms, and Azure Kubernetes Service (AKS), with ground truth data for accurate assessment. Agent performance is measured through both final detection results and trajectory-based rewards that capture decision-making effectiveness. 
 
 This work demonstrates the potential of AI agents to support labor-intensive aspects of detection engineering. Our comprehensive evaluation of 16 frontier models shows that Claude Opus 4.6 (High) achieves the highest overall reward (0.637), followed by Claude Opus 4.5 (0.624) and the GPT-5 family. An ablation study confirms that CTI-specific tools significantly improve agent performance, a variance analysis across repeated runs demonstrates result stability. Finally, a memory augmentation study shows that seeded context can close 33\% of the performance gap between smaller and larger models.

\end{abstract}

\section{Introduction}

Detection engineering is the process of creating rules and queries to identify malicious activity in security telemetry. It remains a cornerstone of blue team operations. There are two primary challenges that make it difficult: (1) enterprise environments are heterogeneous, spanning endpoints, cloud infrastructure, and containerized workloads, and (2) analysts must interpret threat intelligence, understand diverse data schemas, and iteratively refine detection logic.

Existing cybersecurity benchmarks primarily evaluate parametric knowledge or isolated subtasks such as rule synthesis, TTP classification, and threat actor attribution. None address the end-to-end detection engineering workflow, where an analyst must interpret threat intelligence, explore heterogeneous telemetry, iteratively construct queries, and produce validated detection rules. This gap leaves the field without a principled way to measure whether AI agents can perform the complete analytical pipeline that security teams rely on daily.

We introduce CTI-REALM (Cyber Threat Real World Evaluation and LLM Benchmarking), a benchmark that evaluates AI agents on end-to-end detection engineering in realistic settings. The benchmark is grounded in authentic telemetry from real attack emulations executed on sandboxed Azure infrastructure, where agents use specialized tools to analyze CTI reports and construct queries to produce detection rules. We introduce two benchmarks : CTI-REALM-25 and CTI-REALM-50, with task difficulty ranging from atomic attacks to complex multi-step intrusion scenarios. The benchmark evaluates both final detection quality and decision-making processes through trajectory-based rewards.

Our evaluation of 16 frontier models reveals significant performance variation, with Claude Opus 4.6 (High) achieving the highest reward (0.637), followed by Claude Opus 4.5 (0.624) and GPT-5 variants.

Our contributions are as follows:
\begin{itemize}
    \item A realistic evaluation environment with authentic attack telemetry, containerized sandbox, queryable logs across Linux/AKS/Cloud, and reference databases for MITRE techniques and Sigma rules.
    \item A trajectory-based evaluation framework combining deterministic checkpoints with LLM-as-judge assessment. The fine-grained reward structure at intermediate checkpoints provides training signals suitable for reinforcement learning approaches, enabling hill climbing and RL-based policy optimization for detection engineering.
    \item Comprehensive ground-truth-annotated benchmark dataset spanning three platforms at varying difficulty levels.
    \item Empirical evaluation of 16 frontier models with ablation and variance analyses.
\end{itemize}

\section{Background and Previous Work}

Detection engineering has become a focal point for LLM-based automation in cybersecurity. This section reviews relevant work around LLM-driven detection rule generation, threat intelligence, cybersecurity benchmarks, and situates CTI-REALM in this landscape.

\subsection{LLM-Based Detection Rule Generation}

Several systems have explored automating detection rule creation. LLMCloudHunter [1] generates Sigma rules from unstructured CTI reports with multimodal analysis, achieving 92 percent precision for API call extraction. IntelEX [2] extracts attack-level TTPs using in-context learning, while SigmaGen [3] fine-tunes LLMs specifically for Sigma rule generation. RuleGenie [4] addresses rule set optimization to reduce redundancy, and Sublime Security's ADÉ [5] introduced agentic workflows for email threat detection. Bertiger et al.~[15] present an evaluation framework for LLM-generated detection rules, comparing automated outputs against human-written rule corpora. However, these systems focus on isolated rule synthesis or evaluation rather than end-to-end detection engineering workflows.

\subsection{Threat Intelligence Extraction}
 
Upstream from rule generation, research has addressed extracting structured intelligence from CTI sources. Rule-ATT\&CK Mapper [6] uses multi-stage LLM pipelines to map SIEM rules to MITRE ATT\&CK techniques, while comparative studies [7] have evaluated encoder-based models versus decoder-based LLMs with RAG for TTP classification. These approaches demonstrate LLM potential but rely on static evaluation without iterative refinement.

\subsection{Cybersecurity Benchmarks}
 
CTIBench [8] evaluates LLMs on CTI tasks including root cause mapping and threat actor attribution under closed-book settings. AthenaBench [9] extends this with dynamic data from live CTI sources, while ExCyTIn-Bench [10] evaluates agents on threat investigation using Microsoft Sentinel logs. CAIBench [11] revealed that models achieving ~70 percent on knowledge benchmarks degrade to 20-40 percent on multi-step adversarial scenarios, highlighting gaps between knowledge and capability. CyberSOCEval~[16] benchmarks LLMs on malware analysis and threat intelligence reasoning within a SOC context, finding that reasoning models do not achieve the same performance boost as in coding and math domains. CTIArena~[18] evaluates LLMs across heterogeneous CTI tasks under both closed-book and knowledge-augmented settings, demonstrating significant gains from retrieval augmentation. These benchmarks primarily test parametric knowledge rather than tool-augmented detection engineering.

\subsection{Attack Simulation Frameworks}

CTI-REALM leverages established adversary emulation tools as inspiration for ground truth generation. Atomic Red Team~[12] provides a library of small, portable tests mapped to the MITRE ATT\&CK framework, enabling reproducible simulation of individual attack techniques. Several of our simulations adapt Atomic Red Team test cases to produce realistic telemetry in controlled environments.

CTI-REALM differs from prior work by: (1) evaluating complete detection engineering workflows rather than isolated rule synthesis; (2) providing realistic tools for active data exploration rather than testing parametric knowledge; (3) introducing trajectory-based evaluation capturing decision-making processes; and (4) grounding evaluation in authentic telemetry from attacks executed on real infrastructure across Linux, AKS, and Azure cloud environments.

\section{Benchmark Design and Methodology}

CTI-REALM is designed to evaluate AI agents on the complete detection engineering workflow. This extends from interpreting threat intelligence to producing validated detection rules. The benchmark presents objectives derived from real-world CTI reports and requires agents to analyze relevant threat intelligence, identify appropriate data sources, iteratively construct and test queries, and produce both Sigma rules and executable KQL queries.

\subsection{Task Formulation}

The core task in CTI-REALM mirrors the workflow of SOC analysts. Given a CTI report describing an attack, the goal is to develop detection logic that identifies the threat in production environments. Each task instance presents an agent with:

\begin{itemize}\setlength{\itemsep}{1pt}
    \item \textbf{Input}: CTI references from public detection repositories and threat research blogs, plus telemetry logs from the attack emulation.
    \item \textbf{Output}: Detection rules in Sigma and KQL\footnote{Kusto Query Language (KQL) is Microsoft's query language for Azure Data Explorer, Microsoft Sentinel, and other Azure services. See \url{https://learn.microsoft.com/en-us/kusto/query/}.} formats that identify the described attack patterns while minimizing false positives.
    \item \textbf{Constraints}: Agents operate within the containerized environment using provided tools only. They cannot modify telemetry or access external resources.
\end{itemize}

Figure~\ref{fig:task_examples} illustrates two representative tasks.

\begin{figure}[h]
\begin{tcolorbox}[colback=green!3, colframe=green!40!black, title={\textbf{Example 1: Linux Endpoint Detection (Easy)}}, fonttitle=\small, top=2pt, bottom=2pt]
\scriptsize
\textbf{Task:} \textit{Write a detection rule to identify scheduled task creation using at command by monitoring for shell commands piped to at, indicating attempts to establish persistence or delayed execution of malicious commands.}\\[2pt]
\textbf{Ground Truth ---} \textbf{MITRE:} T1053 (Scheduled Task/Job) \quad \textbf{Sources:} DeviceProcessEvents\\
\textbf{Fields:} \texttt{filename} matches \texttt{at}, \texttt{processcommandline} matches pipe-to-at patterns
\end{tcolorbox}

\vspace{-6pt}

\begin{tcolorbox}[colback=red!3, colframe=red!40!black, title={\textbf{Example 2: Cloud Multi-Step Attack (Hard)}}, fonttitle=\small, top=2pt, bottom=2pt]
\scriptsize
\textbf{Task:} \textit{Write a detection rule to identify the creation of restricted management Administrative Units where privileged administrators create organizational containers with management restriction properties enabled, then add user accounts as members. Monitor for AU creation through directory management APIs, followed by member addition activities.}\\[2pt]
\textbf{Ground Truth ---} \textbf{MITRE:} T1078, T1069, T1098, T1484, T1136 \quad \textbf{Sources:} SigninLogs, MicrosoftGraphActivityLogs, AuditLogs\\
\textbf{Fields:} \texttt{RequestUri} matches \texttt{/directory/administrativeUnits}, \texttt{OperationName} $\in$ \{Add administrative unit, Add member to restricted management administrative unit\}
\end{tcolorbox}
\caption{Representative CTI-REALM tasks spanning the difficulty spectrum.}
\label{fig:task_examples}
\end{figure}

\subsection{Simulation and Data Collection}

CTI-REALM's dataset comprises attack simulations executed across three distinct environments: Linux endpoints, Azure Kubernetes Service (AKS), and Azure cloud infrastructure. These simulations were derived from 37 publicly available CTI reports and detection references sourced from Microsoft Security, Datadog Security Labs, Palo Alto Networks and Splunk Security Content. The selection criteria prioritized (1) variation in attack complexity, (2) feasibility of accurate recreation, (3) viability for detection rule development, and (4) representation across difficulty levels.

\textbf{Difficulty and Dataset Composition}: Simulations span three difficulty tiers - easy (atomic single-step attacks), medium (multi-step sequences), and hard (complex attack chains requiring cross-source correlation). Cloud simulations are exclusively hard, reflecting real-world APT campaign complexity. From the full simulation pool, we construct two stratified evaluation sets: CTI-REALM-25 (12 Linux, 9 AKS, 4 Cloud) and CTI-REALM-50 (25 Linux, 17 AKS, 8 Cloud). CTI-REALM-25 is a subset of CTI-REALM-50. The smaller set enables rapid iteration and development testing with lower cost, while CTI-REALM-50 provides a more robust evaluation surface with greater coverage across difficulty levels and platform categories.

\textbf{Attack Execution and Log Collection}: An important architectural distinction in CTI-REALM is the separation between attack execution and agent evaluation environments. Simulations are executed on actual infrastructure, including Linux endpoints, AKS clusters, and other Azure cloud infrastructure hosted on an isolated sandboxed Azure tenant to maintain security and traceability. Realistic telemetry is collected through the Azure Monitor Agent and MDE (Microsoft Defender for Endpoint). Prior to evaluation, logs undergo cleaning and anonymization to remove personally identifiable information (PII). We also sanitize sensitive infrastructure details and prevent benchmark contamination by removing resource group names, key vault identifiers etc. The cleaned logs are then transferred to the containerized evaluation environment where agents perform analysis.

\subsection{Environment Architecture}

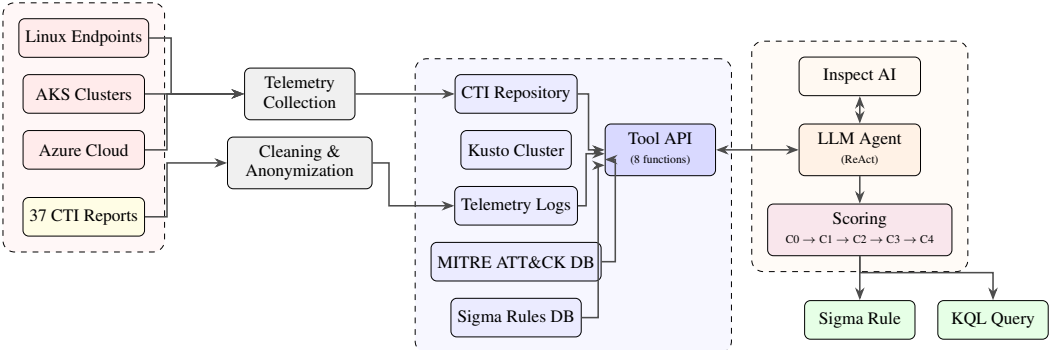
\begin{figure}[t]
\centering
\resizebox{\textwidth}{!}{%
\begin{tikzpicture}[
    >=Stealth,
    node distance=0.4cm,
    box/.style={draw, rounded corners=3pt, minimum height=0.7cm, text centered, font=\small},
    infra/.style={box, fill=red!8, minimum width=2.2cm},
    container/.style={box, fill=blue!8, minimum width=2.0cm},
    agent/.style={box, fill=orange!10, minimum width=2.2cm},
    output/.style={box, fill=green!10, minimum width=2.0cm},
    group/.style={draw, dashed, rounded corners=6pt, inner sep=8pt},
    arrowstyle/.style={->, thick, color=black!70},
    label/.style={font=\footnotesize\bfseries, text=black!70},
]

\node[infra] (linux) {Linux Endpoints};
\node[infra, below=0.3cm of linux] (aks) {AKS Clusters};
\node[infra, below=0.3cm of aks] (cloud) {Azure Cloud};

\node[box, fill=yellow!12, minimum width=2.2cm, below=0.5cm of cloud] (cti_src) {37 CTI Reports};

\begin{scope}[on background layer]
\node[group, fill=red!3, fit=(linux)(aks)(cloud)(cti_src), label={[label]above:Sandboxed Azure Tenant}] (infra_group) {};
\end{scope}

\node[box, fill=gray!12, minimum width=2.0cm, right=1.8cm of aks] (collect) {\begin{tabular}{c}Telemetry\\Collection\end{tabular}};
\node[box, fill=gray!12, minimum width=2.0cm, below=0.3cm of collect] (clean) {\begin{tabular}{c}Cleaning \&\\Anonymization\end{tabular}};

\node[container, right=1.8cm of collect] (cti_repo) {CTI Repository};
\node[container, below=0.3cm of cti_repo] (kusto) {Kusto Cluster};
\node[container, below=0.3cm of kusto] (logs) {Telemetry Logs};
\node[container, below=0.3cm of logs] (mitre) {MITRE ATT\&CK DB};
\node[container, below=0.3cm of mitre] (sigma) {Sigma Rules DB};

\node[box, fill=blue!15, minimum width=2.0cm, right=0.6cm of kusto] (toolapi) {\begin{tabular}{c}Tool API\\{\tiny(8 functions)}\end{tabular}};

\begin{scope}[on background layer]
\node[group, fill=blue!3, fit=(cti_repo)(kusto)(logs)(mitre)(sigma)(toolapi), label={[label]above:Docker Container}] (container_group) {};
\end{scope}

\node[agent, right=1.5cm of toolapi] (llm) {\begin{tabular}{c}LLM Agent\\{\tiny(ReAct)}\end{tabular}};
\node[box, fill=orange!5, minimum width=2.2cm, above=0.5cm of llm] (inspect) {Inspect AI};

\node[box, fill=purple!10, minimum width=2.2cm, below=0.5cm of llm] (checkpoints) {\begin{tabular}{c}Scoring\\{\tiny C0 $\rightarrow$ C1 $\rightarrow$ C2 $\rightarrow$ C3 $\rightarrow$ C4}\end{tabular}};

\begin{scope}[on background layer]
\node[group, fill=orange!3, fit=(inspect)(llm)(checkpoints), label={[label]above:Agent \& Evaluation}] (agent_group) {};
\end{scope}

\node[output, below=0.8cm of checkpoints] (sigmarule) {Sigma Rule};
\node[output, right=0.4cm of sigmarule] (kql) {KQL Query};

\draw[arrowstyle] (linux.east) -- ++(0.4,0) |- (collect.west);
\draw[arrowstyle] (aks.east) -- (collect.west);
\draw[arrowstyle] (cloud.east) -- ++(0.4,0) |- (collect.west);
\draw[arrowstyle] (cti_src.east) -- ++(0.4,0) |- (clean.west);

\draw[arrowstyle] (collect.east) -- (cti_repo.west);
\draw[arrowstyle] (clean.east) -- ++(0.3,0) |- (logs.west);

\draw[arrowstyle] (cti_repo.east) -- ++(0.2,0) |- (toolapi.west);
\draw[arrowstyle] (logs.east) -- ++(0.15,0) |- ([yshift=-2pt]toolapi.west);
\draw[arrowstyle] (mitre.east) -- ++(0.25,0) |- ([yshift=-5pt]toolapi.west);
\draw[arrowstyle] (sigma.east) -- ++(0.3,0) |- ([yshift=-8pt]toolapi.west);

\draw[arrowstyle, <->] (toolapi.east) -- (llm.west);

\draw[arrowstyle, <->] (inspect.south) -- (llm.north);

\draw[arrowstyle] (llm.south) -- (checkpoints.north);

\draw[arrowstyle] (checkpoints.south) -- ++(0,-.3) -| (sigmarule.north);
\draw[arrowstyle] (checkpoints.south) -- ++(0,-.3) -| (kql.north);

\end{tikzpicture}
}%
\caption{CTI-REALM environment architecture.}
\label{fig:architecture}
\end{figure}

The CTI-REALM evaluation environment is implemented as a containerized Docker system that replicates the operational workspace of security detection engineers. It is integrated with Inspect AI~[14]-a framework for LLM agent evaluation to provide standardized interfaces for task execution and scoring. This architecture provides agents with the necessary tools and data to analyze threat intelligence and develop detection rules while maintaining isolation and reproducibility. Figure~\ref{fig:architecture} illustrates the end-to-end system.

\textbf{Environment Components} : The container hosts several key resources : 

\begin{itemize}
    \item \textbf{CTI repository} : 37 source reports from Microsoft Security, Datadog Security Labs, Palo Alto Networks, and Splunk Security Content that served as the basis for attack simulations
    \item \textbf{Kusto Cluster} :  A query engine enabling agents to execute KQL (Kusto Query Language) queries against telemetry data
    \item \textbf{Telemetry logs} : Multi-source security logs collected from the attack simulations
    \item \textbf{MITRE ATT\&CK Database}~[17] : Techniques and tactic mappings for threat contextualization
    \item \textbf{Sigma rules database} : A reference collection of existing detection rules to prevent duplication and provide context
\end{itemize}

\textbf{Telemetry Data Sources}: The environment contains 12 log sources collected through MDE, Azure Monitor, and Azure AD: endpoint telemetry (\texttt{deviceprocessevents}, \texttt{devicefilevents}), AKS logs (\texttt{aksaudit}, \texttt{aksauditadmin}), Azure cloud operations (\texttt{azureactivity}, \texttt{azurediagnostics}), identity and authentication (\texttt{signinlogs}, \texttt{auditlogs}, \texttt{aadserviceprincipalsigninlogs}), and application-layer logs (\texttt{officeactivity}, \texttt{microsoftgraphactivitylogs}, \texttt{storagebloblogs}). Each source contains 1--3 days of telemetry incorporating both attack-generated events and benign background activity.

\textbf{Agent Tools}: Agents interact with the environment through a structured API providing eight specialized functions, detailed in the appendix. These tools support CTI report retrieval, data exploration, query execution, and threat context mapping.

\subsection{Evaluation Framework}

CTI-REALM frames detection rule development as a sequential decision-making task within a reinforcement learning environment. Rather than treating detection engineering as a single-step generation problem, the benchmark models it as a multi-stage process where agents receive reward signals at intermediate checkpoints corresponding to natural progression points in the analytical workflow.

\textbf{Reinforcement Learning Formulation:} We model each detection task as a Markov Decision Process (MDP)~[21] where:
\begin{itemize}
    \item \textbf{State space} $S$ captures the agent's current understanding (CTI reports retrieved, TTPs identified, data sources explored, query results obtained)
    \item \textbf{Action space} $A$ consists of tool invocations (CTI retrieval, schema inspection, query execution, etc.)
    \item \textbf{Reward function} $R$ provides feedback at five checkpoints throughout the trajectory
\end{itemize}

The total reward for a task is computed as:

\begin{equation}
    R_{\text{total}} = \sum_{i \in \{C0, C1, C2, C3, C4\}} w_i \cdot r_i
\end{equation}

where $w_i$ represents the weight assigned to checkpoint $i$ (with $\sum w_i = 1$) and $r_i \in [0,1]$ is the normalized reward achieved at that checkpoint, yielding $R_{\text{total}} \in [0,1]$. This decomposes into checkpoint and ground truth components:

\begin{equation}
    R_{\text{total}} = R_{\text{checkpoint}} + R_{\text{ground truth}}
\end{equation}

where:
\begin{itemize}
    \item $R_{\text{checkpoint}} = \sum_{i \in \{C0, C1, C2, C3\}} w_i \cdot r_i$ (35\% total weight)
    \item $R_{\text{ground truth}} = w_{C4} \cdot r_{C4}$ (65\% total weight)
\end{itemize}

\textbf{Checkpoint Reward Signals:} Each checkpoint evaluates a distinct aspect of the detection development process:

\begin{itemize}
    \item \textbf{C0 - CTI Report Analysis} ($w_{C0} = 0.125$, LLM-as-judge): Rewards correct identification and retrieval of relevant threat intelligence reports
    \item \textbf{C1 - Threat Context} ($w_{C1} = 0.075$, Jaccard similarity): Rewards accurate extraction and mapping of MITRE ATT\&CK techniques
    \item \textbf{C2 - Data Exploration} ($w_{C2} = 0.10$, Jaccard similarity): Rewards identification of relevant telemetry sources through schema exploration
    \item \textbf{C3 - Query Execution} ($w_{C3} = 0.05$, binary): Rewards iterative query refinement behavior ($\geq$2 successful queries)
    \item \textbf{C4 - Detection Quality} ($w_{C4} = 0.65$, F1 + LLM-as-judge): Rewards detection rule effectiveness (KQL correctness via F1-score, Sigma rule quality via judge)
\end{itemize}

This formulation serves dual purposes: (1) enabling comparative evaluation of reasoning capabilities across agent architectures, and (2) providing training signals for reinforcement learning approaches that leverage intermediate rewards to learn effective security analysis policies.

\textbf{Evaluation Methodologies:} The framework uses both deterministic and non-deterministic evaluation. The deterministic section includes tool usage verification (for C0 and C3), Jaccard similarity for comparing identified TTPs and data sources against ground truth (C1, C2). Finally, for detection accuracy we use regexes and F1-score. Non-deterministic evaluation utilizes GPT-5-Mini as an LLM-as-a-judge~[19] for assessing report relevance and detection rule quality. To validate judge reliability, a sample of LLM-as-judge outputs was manually reviewed by security researchers, confirming alignment between automated rewards and expert assessment. Full details on ground truth structure, judge prompts, and scoring calibration are provided in Appendix~\ref{app:eval_details} and Appendix~\ref{app:judge}.

\section{Experimental Setup}

We evaluate agent performance on CTI-REALM using a ReAct agent framework implemented on the Inspect AI~[14] platform. This section describes the models tested, agent architecture, and experimental configurations.

\textbf{Models Evaluated:} We assess 16 model configurations: three Anthropic Claude models (Opus 4.6 at the default high reasoning effort~[22], Opus 4.5, Sonnet 4.5), nine OpenAI GPT-5 family variants (GPT-5, 5.1, 5.2 each at high/medium/low reasoning effort), GPT-5-Mini, GPT-4.1, and two reasoning models (O3, O4-Mini). The reasoning effort levels control the computational budget for extended thinking, enabling analysis of how reasoning depth affects detection performance.

\textbf{Agent Architecture:} All evaluations employ a ReAct~[13] agent that interleaves reasoning traces with tool invocations, mirroring how security analysts alternate between analytical thinking and practical investigation. We use a single agent architecture to isolate model capability differences under controlled conditions; comparing multiple agent frameworks (e.g., plan-and-execute, tree-of-thought) is left to future work. Agents are permitted a maximum of 70 messages per task. Each model is evaluated on CTI-REALM-50 using identical configurations.

\textbf{Evaluation Protocol:} We conduct evaluation in four phases to balance comprehensive model coverage with statistical rigor. In the initial phase, all 16 model configurations are evaluated on CTI-REALM-50 to establish baseline rankings. In the second phase, we select the top five models for repeated evaluation (three epochs) on CTI-REALM-25 to assess variance and establish confidence intervals. In the third phase, we conduct an ablation study on the same top five models using CTI-REALM-25 with minimal tools (removing CTI-specific capabilities) to measure the impact of tool augmentation. In the fourth phase, we investigate memory augmentation by comparing GPT-5-Mini with and without seeded memory context against the full GPT-5 model at multiple reasoning effort levels. This approach enables broad comparison across model families.

\section{Results}

We present results across four experimental setups: comprehensive model comparison on CTI-REALM-50 (Sections~\ref{sec:overall}--\ref{sec:efficiency}), variance analysis with repeated runs (Section~\ref{sec:variance}), an ablation study examining the impact of CTI-specific tools (Section~\ref{sec:ablation}), and a memory augmentation study (Section~\ref{sec:memory}).

\subsection{Overall Model Performance}
\label{sec:overall}

Table~\ref{tab:main_results} presents the complete results for all 16 model configurations evaluated on CTI-REALM-50. Claude Opus 4.6 (High) achieves the highest normalized reward of 0.637 ($\pm$0.037 SE), followed closely by Claude Opus 4.5 at 0.624 ($\pm$0.034). Anthropic models occupy the top three positions, with Claude Sonnet 4.5 at 0.587. Among OpenAI models, GPT-5 (Med) leads at 0.572, closely followed by GPT-5.2 (Med) at 0.572.

\begin{figure}[h]
    \centering
    \includegraphics[width=\textwidth]{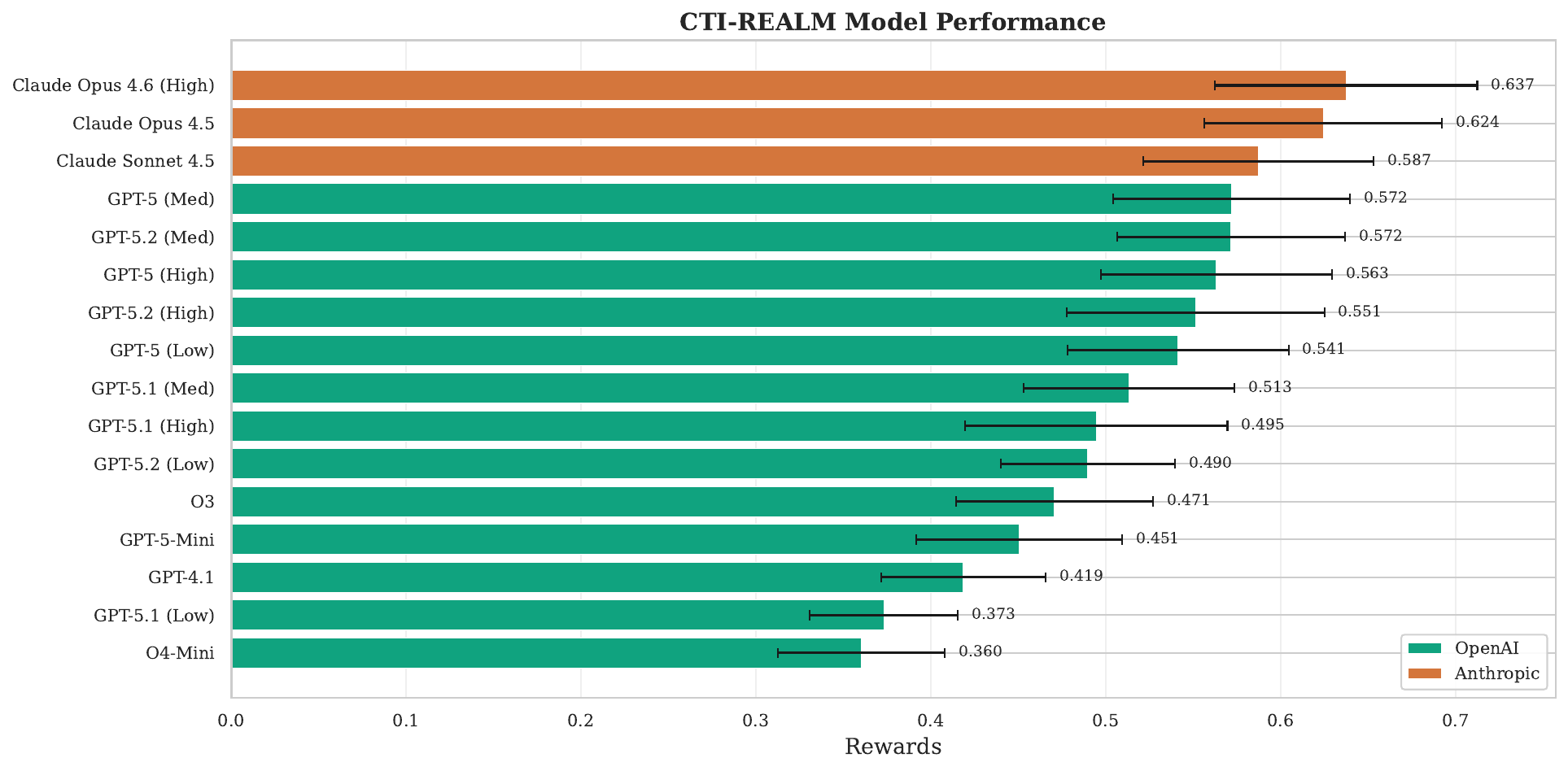}
    \caption{Model performance on CTI-REALM-50, sorted by normalized reward.}
    \label{fig:model_performance}
\end{figure}

\begin{table}[t]
\centering
\caption{CTI-REALM Model Performance Summary. All scores normalized to [0,1]. CI = 95\% confidence interval. Checkpoint Reward + Ground Truth Reward = Rewards.}
\label{tab:main_results}
\resizebox{\textwidth}{!}{%
\begin{tabular}{llcccccccccc}
\toprule
Model & Provider & Rewards & StdErr & 95\% CI & Checkpoint Reward & Ground Truth Reward & C0 & C1 & C2 & C3 & Steps \\
\midrule
Claude Opus 4.6 (High) & Anthropic & 0.6373 & 0.0374 & [0.562, 0.712] & 0.266 & 0.372 & 0.83 & 0.50 & 0.82 & 0.86 & 31 \\
Claude Opus 4.5 & Anthropic & 0.6244 & 0.0338 & [0.556, 0.692] & 0.287 & 0.337 & 0.91 & 0.56 & 0.85 & 0.92 & 32 \\
Claude Sonnet 4.5 & Anthropic & 0.5872 & 0.0328 & [0.521, 0.653] & 0.278 & 0.310 & 0.85 & 0.54 & 0.87 & 0.88 & 37 \\
GPT-5 (Med) & OpenAI & 0.5720 & 0.0337 & [0.504, 0.640] & 0.235 & 0.337 & 0.72 & 0.54 & 0.86 & 0.38 & 32 \\
GPT-5.2 (Med) & OpenAI & 0.5716 & 0.0325 & [0.506, 0.637] & 0.226 & 0.346 & 0.60 & 0.37 & 0.86 & 0.72 & 31 \\
GPT-5 (High) & OpenAI & 0.5633 & 0.0329 & [0.497, 0.629] & 0.230 & 0.333 & 0.66 & 0.55 & 0.86 & 0.40 & 32 \\
GPT-5.2 (High) & OpenAI & 0.5513 & 0.0367 & [0.478, 0.625] & 0.232 & 0.319 & 0.60 & 0.40 & 0.85 & 0.84 & 35 \\
GPT-5 (Low) & OpenAI & 0.5413 & 0.0315 & [0.478, 0.605] & 0.204 & 0.337 & 0.61 & 0.45 & 0.84 & 0.22 & 25 \\
GPT-5.1 (Med) & OpenAI & 0.5133 & 0.0300 & [0.453, 0.574] & 0.244 & 0.269 & 0.72 & 0.62 & 0.87 & 0.42 & 26 \\
GPT-5.1 (High) & OpenAI & 0.4946 & 0.0373 & [0.420, 0.570] & 0.238 & 0.256 & 0.71 & 0.55 & 0.80 & 0.56 & 29 \\
GPT-5.2 (Low) & OpenAI & 0.4898 & 0.0248 & [0.440, 0.540] & 0.221 & 0.269 & 0.57 & 0.49 & 0.88 & 0.50 & 27 \\
O3 & OpenAI & 0.4707 & 0.0281 & [0.414, 0.527] & 0.204 & 0.267 & 0.51 & 0.38 & 0.78 & 0.68 & 33 \\
GPT-5-Mini & OpenAI & 0.4506 & 0.0292 & [0.392, 0.509] & 0.188 & 0.263 & 0.62 & 0.21 & 0.84 & 0.20 & 27 \\
GPT-4.1 & OpenAI & 0.4186 & 0.0235 & [0.371, 0.466] & 0.193 & 0.225 & 0.68 & 0.29 & 0.85 & 0.02 & 21 \\
GPT-5.1 (Low) & OpenAI & 0.3731 & 0.0211 & [0.331, 0.415] & 0.194 & 0.179 & 0.61 & 0.37 & 0.87 & 0.06 & 22 \\
O4-Mini & OpenAI & 0.3602 & 0.0238 & [0.312, 0.408] & 0.167 & 0.193 & 0.48 & 0.23 & 0.80 & 0.20 & 26 \\
\bottomrule
\end{tabular}%
}
\end{table}

The results reveal several notable patterns across the 0.637--0.419 performance range. Anthropic Claude models occupy the top three positions, with a clear separation from the best OpenAI result (GPT-5 Med, 0.572). This advantage appears driven by stronger tool-use and agentic capabilities, with particularly large gaps in query execution (C3) and CTI comprehension (C0), as detailed in Section~\ref{sec:checkpoints}.

Within the GPT-5 family, medium reasoning effort achieves the best results across all three generations (GPT-5, 5.1, 5.2), while high reasoning consistently underperforms medium. This pattern is consistent with overthinking: higher reasoning budgets may cause the agent to over-elaborate on analytical steps or second-guess effective query strategies, a phenomenon observed in other agentic benchmarks where additional reasoning computation yields diminishing or negative returns. The dedicated reasoning models O3 and O4-Mini similarly underperform general-purpose models, reinforcing that extended chain-of-thought reasoning does not directly translate to improved detection engineering capabilities. Of the 120 pairwise comparisons, 71 (59.2\%) are statistically significant ($\alpha = 0.05$), with 34 (28.3\%) remaining significant after Bonferroni correction. Table~\ref{tab:main_results} also reports per-checkpoint rewards: C0 (CTI report analysis), C1 (MITRE technique mapping), C2 (data source exploration), and C3 (query execution rate).

\subsection{Category Performance}
\label{sec:category}

\begin{figure}[h]
    \centering
    \begin{subfigure}[t]{0.49\textwidth}
        \vspace{0pt}
        \centering
        \includegraphics[width=\textwidth,height=6.5cm,keepaspectratio]{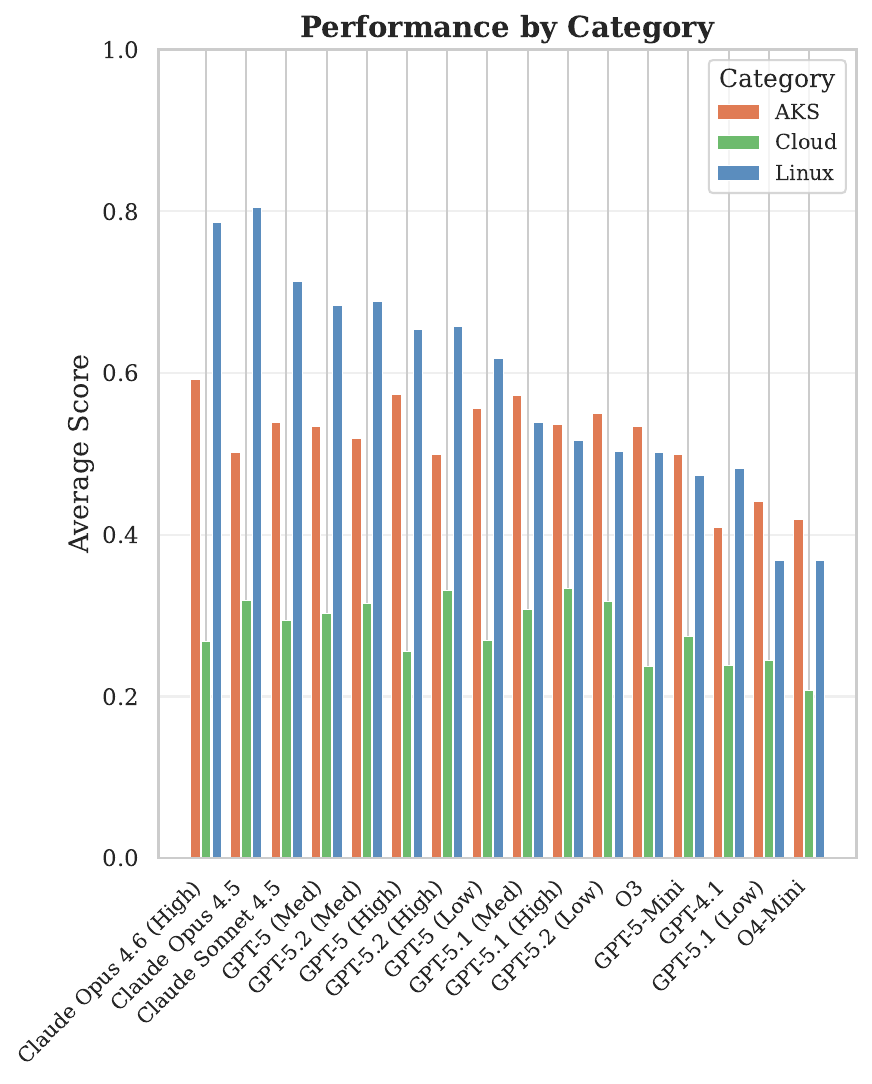}
        \caption{Performance by environment category.}
        \label{fig:category_performance}
    \end{subfigure}%
    \hfill
    \begin{subfigure}[t]{0.49\textwidth}
        \vspace{0pt}
        \centering
        \includegraphics[width=\textwidth,height=6.5cm,keepaspectratio]{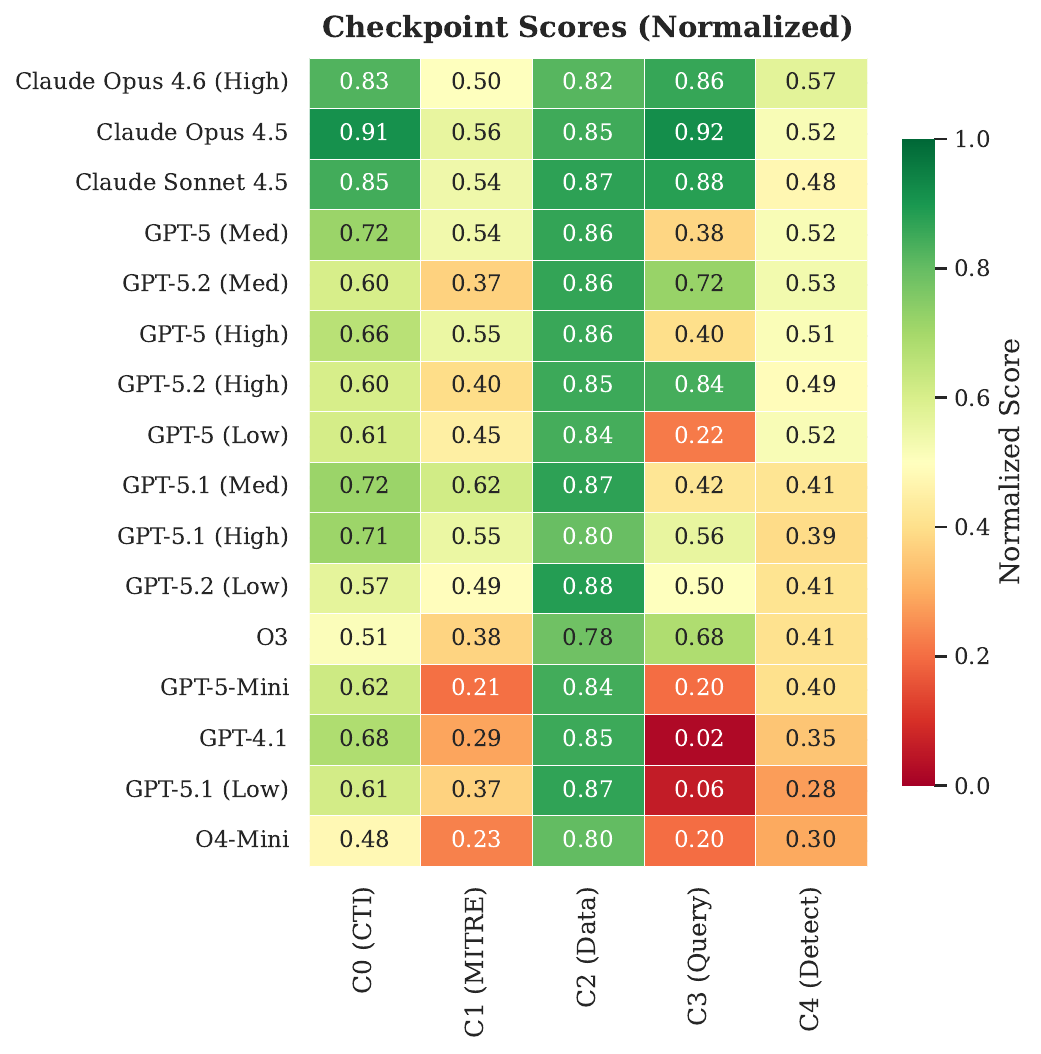}
        \caption{Mean checkpoint rewards (C0--C4, normalized).}
        \label{fig:checkpoint_heatmap}
    \end{subfigure}
    \caption{Category and checkpoint analysis on CTI-REALM-50.}
    \label{fig:category_and_checkpoints}
\end{figure}

Performance varies substantially across platform categories (Figure~\ref{fig:category_performance}). Linux tasks yield the highest average rewards (0.585 across all models), followed by AKS (0.517) and Cloud (0.282). This difficulty gradient reflects the underlying task complexity: Linux simulations include many easy-difficulty atomic attacks with straightforward process-level telemetry. Meanwhile, AKS tasks require understanding container orchestration logs, while Cloud tasks involve multi-step attack chains requiring correlation across disparate data sources (Azure Activity, Sign-in logs, Diagnostics).

The Cloud category, where all 8 simulations are classified as hard difficulty, represents the most realistic APT-style detection scenarios and reveals the steepest performance drop across all models. Even the top-performing Claude Opus 4.6 (High) shows a substantial decline on Cloud tasks compared to Linux, demonstrating that current frontier models struggle with the correlation and contextualization demands of advanced cloud threat detection.

\subsection{Checkpoint Analysis}
\label{sec:checkpoints}

Figure~\ref{fig:checkpoint_heatmap} shows normalized mean rewards for each checkpoint. C3 (query execution) is the most discriminating - Claude models score 0.86--0.92 while most OpenAI models fall below 0.50, with GPT-4.1 at just 0.02. C1 (MITRE mapping) varies widely (0.21--0.56), with Claude Opus 4.5 leading at 0.56, suggesting that strong CTI knowledge does not guarantee downstream performance. C0 (CTI report analysis) separates Claude (0.85--0.91) from most OpenAI models (0.49--0.70), reflecting differences in threat intelligence comprehension that propagate through subsequent stages.

\subsection{Efficiency Analysis}
\label{sec:efficiency}

\begin{figure}[h]
    \centering
    \begin{subfigure}[t]{0.48\textwidth}
        \vspace{0pt}
        \centering
        \includegraphics[width=\textwidth]{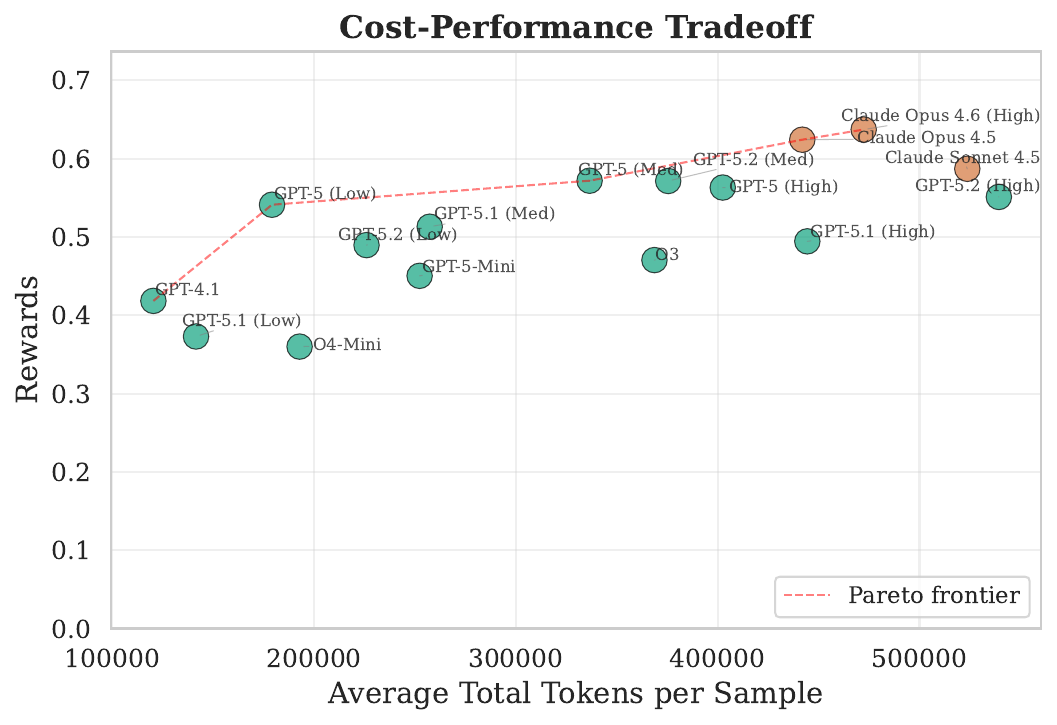}
        \caption{Token usage vs.\ reward with Pareto frontier.}
        \label{fig:cost_performance}
    \end{subfigure}
    \hfill
    \begin{subfigure}[t]{0.48\textwidth}
        \vspace{0pt}
        \centering
        \includegraphics[width=\textwidth]{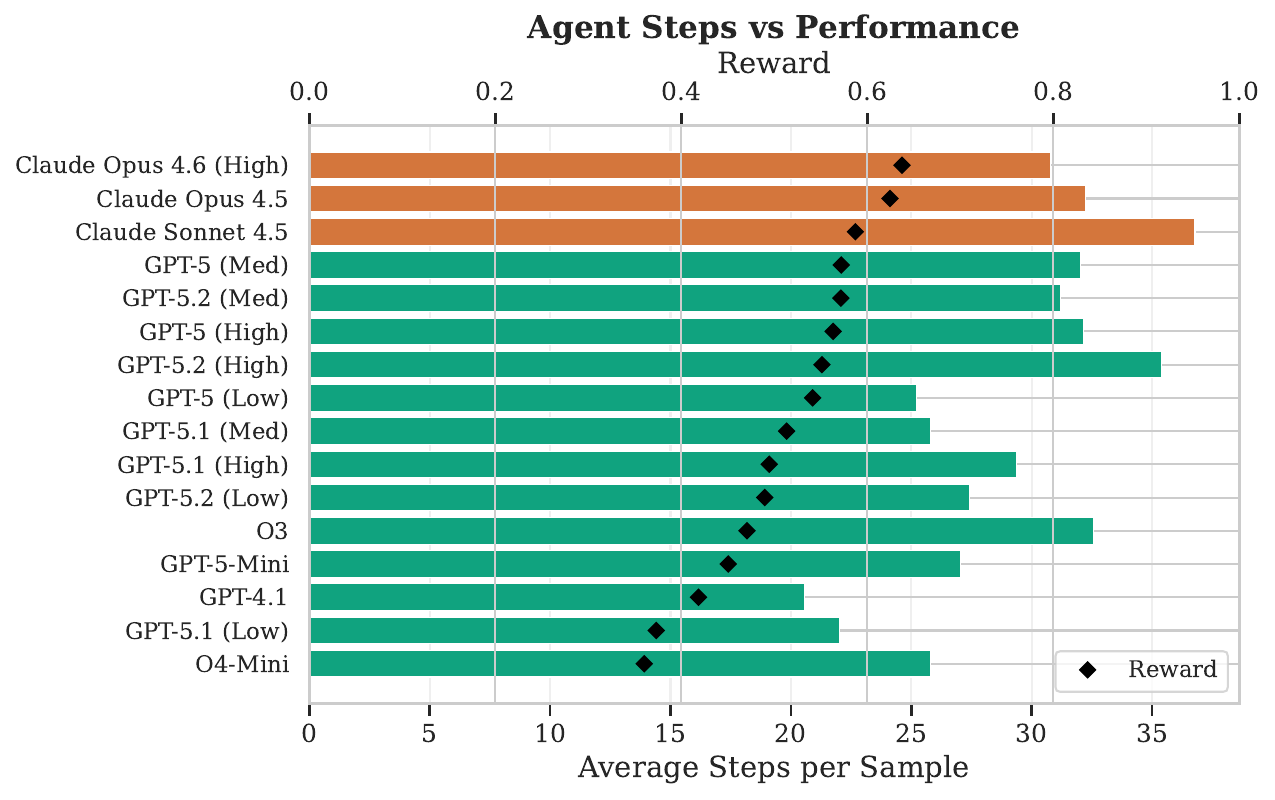}
        \caption{Average steps per sample vs.\ reward.}
        \label{fig:steps_performance}
    \end{subfigure}
    \caption{Cost and interaction efficiency on CTI-REALM-50.}
    \label{fig:efficiency}
\end{figure}

Figure~\ref{fig:cost_performance} illustrates the cost-performance tradeoff using total tokens per sample (a full input/output breakdown is provided in Table~\ref{tab:token_usage}, Appendix~\ref{app:additional}). GPT-4.1 is the most token-efficient at $\sim$120K tokens (reward 0.419), while GPT-5 (Low) offers the best Pareto-frontier balance at $\sim$179K tokens (reward 0.541). The Anthropic models achieve top rewards with moderate consumption ($\sim$442--524K). Step counts (Figure~\ref{fig:steps_performance}) show that top-performing models use 30--37 steps on average versus 20--27 for weaker models, suggesting effective detection engineering requires sustained iterative investigation.

\subsection{Variance and Reproducibility}
\label{sec:variance}

\begin{table}[t]
\centering
\caption{Score variance across repeated evaluation runs (CTI-REALM-25).}
\label{tab:variance}
\begin{tabular}{lccccc}
\toprule
Model & Mean & Std & Min & Max & 95\% CI \\
\midrule
Claude Opus 4.6 (High) & 0.6130 & 0.2626 & 0.0000 & 0.9426 & [0.553, 0.673] \\
Claude Opus 4.5 & 0.6124 & 0.2522 & 0.0000 & 0.9655 & [0.554, 0.670] \\
GPT-5 (Med) & 0.5454 & 0.2159 & 0.2406 & 0.9002 & [0.496, 0.595] \\
GPT-5.2 (Med) & 0.5331 & 0.2254 & 0.0000 & 0.9456 & [0.481, 0.585] \\
GPT-5.1 (Med) & 0.4815 & 0.1958 & 0.2416 & 0.9195 & [0.436, 0.527] \\
\bottomrule
\end{tabular}
\end{table}

To assess reproducibility, we evaluated the top five models on CTI-REALM-25 across three epochs (Table~\ref{tab:variance}, Figure~\ref{fig:variance}). The rankings remain stable across runs: Claude Opus 4.6 (High) (mean 0.613, std 0.263) and Claude Opus 4.5 (0.612, std 0.252) maintain their leading positions. The within-model variance is substantial (standard deviations of 0.196--0.263),and is primarily driven by task difficulty variation.

Notably, GPT-5 (Med) shows the most consistent performance with a minimum reward of 0.241 (no zero-reward samples), while the Claude models occasionally receive zero rewards on the most difficult tasks. This suggests Claude models take higher-risk analytical approaches that either succeed well or fail completely. The full reward distributions are visualized in Figure~\ref{fig:variance} (Appendix~\ref{app:additional}).

\subsection{Ablation: Impact of CTI Tools}
\label{sec:ablation}

To validate that agents genuinely leverage CTI-specific tools rather than relying on parametric knowledge, we conducted an ablation study removing CTI report retrieval and threat context tools from the agent's toolkit. Table~\ref{tab:ablation} summarizes the results.

\begin{table}[t]
\centering
\caption{Ablation study: impact of removing CTI-specific tools. $\Delta$ = Full $-$ Minimal. Per-checkpoint deltas are normalized; C0 excluded (mechanically zeroed without CTI tools).}
\label{tab:ablation}
\resizebox{\textwidth}{!}{%
\begin{tabular}{lcccccccc}
\toprule
Model & Full Score & Minimal Score & $\Delta$ & Cohen's $d$ & $\Delta$C1 & $\Delta$C2 & $\Delta$C3 & $\Delta$C4 \\
\midrule
Claude Opus 4.6 (High) & 0.6130 & 0.5358 & +0.0772 & 0.30 & -0.17 & -0.13 & -0.12 & +0.01 \\
Claude Opus 4.5 & 0.6124 & 0.4624 & +0.1500 & 0.59 & -0.06 & +0.04 & +0.03 & +0.06 \\
GPT-5 (Med) & 0.5454 & 0.4120 & +0.1335 & 0.64 & -0.07 & -0.02 & +0.16 & +0.07 \\
GPT-5.2 (Med) & 0.5331 & 0.4555 & +0.0776 & 0.35 & -0.14 & -0.03 & +0.04 & +0.02 \\
GPT-5.1 (Med) & 0.4815 & 0.3641 & +0.1174 & 0.61 & -0.05 & +0.02 & -0.04 & +0.04 \\
\bottomrule
\end{tabular}%
}
\end{table}

All five models show performance degradation without CTI tools. Claude Opus 4.5 shows the largest absolute drop ($\Delta = -0.150$), falling from 0.612 to 0.462. GPT-5 (Med) drops by 0.134, and GPT-5.1 (Med) by 0.117. Even the top model, Claude Opus 4.6 (High), drops by 0.077.

Since removing CTI tools mechanically zeroes C0, we exclude it from the per-checkpoint comparison. The right side of Table~\ref{tab:ablation} shows the remaining deltas (positive = full toolkit better). C1 (MITRE mapping) is negative for all models, indicating agents compensate by calling \texttt{search\_mitre\_techniques} more aggressively without CTI context. C3 (query execution) is positive for three of five models, suggesting agents struggle to construct effective queries without CTI guidance. C4 (detection quality) is non-negative across all models, confirming that the overall reward drop is driven by degraded detection rule quality. CTI context is essential for producing high-quality detections, even if intermediate checkpoints can be reached through alternative paths.

\subsection{Memory Augmentation: Closing the Gap with Larger Models}
\label{sec:memory}

We investigate whether smaller models can be augmented to approach larger model performance by providing GPT-5-Mini with seeded memory context. The seeded content consists of human-authored domain guidance: a structured detection engineering workflow, tool-specific usage tips (e.g., KQL query patterns, common debugging strategies), and template patterns for Sigma rules and KQL queries (see Appendix~\ref{app:seed_memory} for examples). Importantly, this is expert-distilled knowledge rather than extracted model trajectories, so the augmentation does not constitute model-to-model distillation. Table~\ref{tab:memory} compares GPT-5-Mini without memory (0.371), with memory (0.432), and GPT-5 at three reasoning levels as upper bounds.

Memory augmentation closes 33\% of the 0.184-point gap to GPT-5 (Med). The improvement is concentrated in knowledge-dependent checkpoints: C1 (MITRE mapping) doubles from 0.22 to 0.44, and the ground truth reward improves by 19\%. However, C3 (query execution) remains unchanged at 0.12, suggesting query construction is an inherent model capability rather than a knowledge gap. The remaining gap to GPT-5 lies primarily in C3 (GPT-5 Med achieves 0.40) and detection quality, indicating the larger model's advantage is in iterative query refinement rather than domain knowledge.

\begin{table}[t]
\centering
\caption{Memory augmentation study results. Seeded = with memory context.}
\label{tab:memory}
\resizebox{\textwidth}{!}{%
\begin{tabular}{lcccccccc}
\toprule
Model & Score & Checkpoint Reward & Ground Truth Reward & C0 & C1 & C2 & C3 & C4 \\
\midrule
GPT-5 (Med) & 0.5556 & 0.229 & 0.326 & 0.72 & 0.54 & 0.80 & 0.40 & 0.50 \\
GPT-5 (Low) & 0.5478 & 0.217 & 0.331 & 0.64 & 0.60 & 0.81 & 0.20 & 0.51 \\
GPT-5 (High) & 0.5348 & 0.249 & 0.286 & 0.73 & 0.58 & 0.83 & 0.64 & 0.44 \\
GPT-5-Mini (Memory) & 0.4324 & 0.207 & 0.225 & 0.66 & 0.44 & 0.86 & 0.12 & 0.35 \\
GPT-5-Mini & 0.3714 & 0.182 & 0.189 & 0.66 & 0.22 & 0.78 & 0.12 & 0.29 \\
\bottomrule
\end{tabular}%
}
\end{table}

\section{Use Cases}

CTI-REALM serves multiple purposes beyond model comparison. Security teams can use it for \textbf{model selection}, identifying which models best suit operational needs. Our results suggest Claude Opus for quality, GPT-5 (Low) for cost-efficiency. The checkpoint-based evaluation enables \textbf{capability gap analysis}: a model scoring high on C2 but low on C3 would benefit from query construction improvements rather than additional CTI context. The ablation and memory studies provide a framework for \textbf{tool and augmentation design}, measuring whether new tools or retrieval strategies meaningfully improve outcomes. Finally, the checkpoint reward structure provides \textbf{training signals for RL-based approaches} to detection engineering.

\section{Limitations}

\textbf{(1) No full SOC replication.} CTI-REALM cannot replicate a production-scale SOC environment, so it does not evaluate query optimization against real telemetry volumes or long-baseline anomaly detection over extended time windows.

\textbf{(2) Azure-centric.} The benchmark targets Azure-native telemetry (MDE, Azure Monitor, Azure AD) and KQL; results may not generalize to AWS/GCP environments or alternative SIEM query languages.

\section{Discussion and Conclusion}

We introduced CTI-REALM, a benchmark for evaluating AI agents on realistic detection engineering workflows. Our evaluation of 16 frontier models reveals that medium reasoning effort consistently outperforms high and low settings. Anthropic Claude models lead overall (Opus 4.6 High: 0.637), and CTI tools provide genuine augmentation (0.019--0.038 point improvements) that propagates through the entire workflow. We also saw that memory augmentation closes 33\% of the gap between GPT-5-Mini and GPT-5 through improved MITRE mapping, while query construction remains an inherent model capability. Cloud-based multi-step detection remains the most significant challenge (0.282 vs 0.585 for Linux). Future work includes expanding platform coverage and data size, incorporating more tooling systems and exploring different types of agentic harnesses.

\section*{Acknowledgements}
The authors would like to thank the Microsoft Security AI Benchmarking and Evaluation team and the broader Microsoft Security (MSEC) organization for their support and contributions to this work.

\newpage
\section*{References}

\medskip

{
\small

\begin{enumerate}
    \item Adar, E., \& Dankwa, R. (2024). LLMCloudHunter: Harnessing LLMs for Automated KQL-Based Cloud Threat Hunting and Sigma Rule Generation. \textit{arXiv preprint arXiv:2407.05099}.

    \item Mitra, S., et al. (2024). IntelEX: A LLM-driven Attack-level Threat Intelligence Extraction Framework. \textit{arXiv preprint arXiv:2407.11191}.

    \item Mukherjee, A., et al. (2024). SigmaGen: Automated Generation of Sigma Rules from Threat Reports. \textit{Proceedings of the AAAI Conference on Artificial Intelligence}.

    \item Gao, P., et al. (2024). RuleGenie: Optimizing Detection Rule Sets with Large Language Models. \textit{IEEE Symposium on Security and Privacy}.

    \item Hall, J. (2024). ADÉ: Agentic Detection Engineering for Email Security. Sublime Security Technical Report.

    \item Al-Shaer, R., et al. (2024). Rule-ATT\&CK Mapper: Mapping SIEM Rules to MITRE ATT\&CK Using Multi-Stage LLM Pipelines. \textit{Digital Threats: Research and Practice}.

    \item Orbinato, V., et al. (2024). Comparative Study on LLMs for TTP Classification: Encoder-Based vs.\ Decoder-Based Models with RAG. \textit{ACM CCS Workshop on AI Security}.

    \item Alam, M., et al. (2024). CTIBench: A Benchmark for Evaluating LLMs in Cyber Threat Intelligence. \textit{arXiv preprint arXiv:2406.07599}.

    \item Gylling, A., et al. (2024). AthenaBench: Evaluating LLMs on Cyber Threat Intelligence with Dynamic Data. \textit{Conference on Applied Machine Learning in Information Security (CAMLIS)}.

    \item Tundis, A., et al. (2025). ExCyTIn-Bench: An Extensible Cyber Threat Investigation Benchmark. \textit{arXiv preprint arXiv:2502.13379}.

    \item Sanz-G\'{o}mez, M., Mayoral-Vilches, V., et al. (2025). Cybersecurity AI Benchmark (CAIBench): A Meta-Benchmark for Evaluating Cybersecurity AI Agents. \textit{arXiv preprint arXiv:2510.24317}.

    \item Red Canary. (2024). Atomic Red Team. \url{https://github.com/redcanaryco/atomic-red-team}.

    \item Yao, S., et al. (2023). ReAct: Synergizing Reasoning and Acting in Language Models. \textit{International Conference on Learning Representations (ICLR)}. arXiv:2210.03629.

    \item UK AI Safety Institute. (2024). Inspect: A Framework for Large Language Model Evaluations. \url{https://inspect.ai-safety-institute.org.uk/}.

    \item Bertiger, A., et al. (2025). Evaluating LLM Generated Detection Rules in Cybersecurity. \textit{Conference on Applied Machine Learning in Information Security (CAMLIS)}. arXiv:2509.16749.

    \item Deason, L., et al. (2025). CyberSOCEval: Benchmarking LLMs Capabilities for Malware Analysis and Threat Intelligence Reasoning. \textit{CyberSecEval 4}. arXiv:2509.20166.

    \item Strom, B.E., et al. (2018). MITRE ATT\&CK: Design and Philosophy. \textit{MITRE Technical Report MTR180312}.

    \item Cheng, Y., et al. (2025). CTIArena: Benchmarking LLM Knowledge and Reasoning Across Heterogeneous Cyber Threat Intelligence. \textit{arXiv preprint arXiv:2510.11974}.

    \item Zheng, L., et al. (2024). Judging LLM-as-a-Judge with MT-Bench and Chatbot Arena. \textit{Advances in Neural Information Processing Systems (NeurIPS)}. arXiv:2306.05685.

    \item Jimenez, C.E., et al. (2024). SWE-bench: Can Language Models Resolve Real-World GitHub Issues? \textit{International Conference on Learning Representations (ICLR)}. arXiv:2310.06770.

    \item Sutton, R.S. and Barto, A.G. (2018). \textit{Reinforcement Learning: An Introduction}. MIT Press, 2nd edition.

    \item Anthropic. (2025). Adaptive Thinking (Extended Thinking). \url{https://platform.claude.com/docs/en/build-with-claude/adaptive-thinking}.
\end{enumerate}

}

\newpage

\appendix

\section{Additional Results}
\label{app:additional}

\begin{figure}[h]
    \centering
    \includegraphics[width=0.55\textwidth]{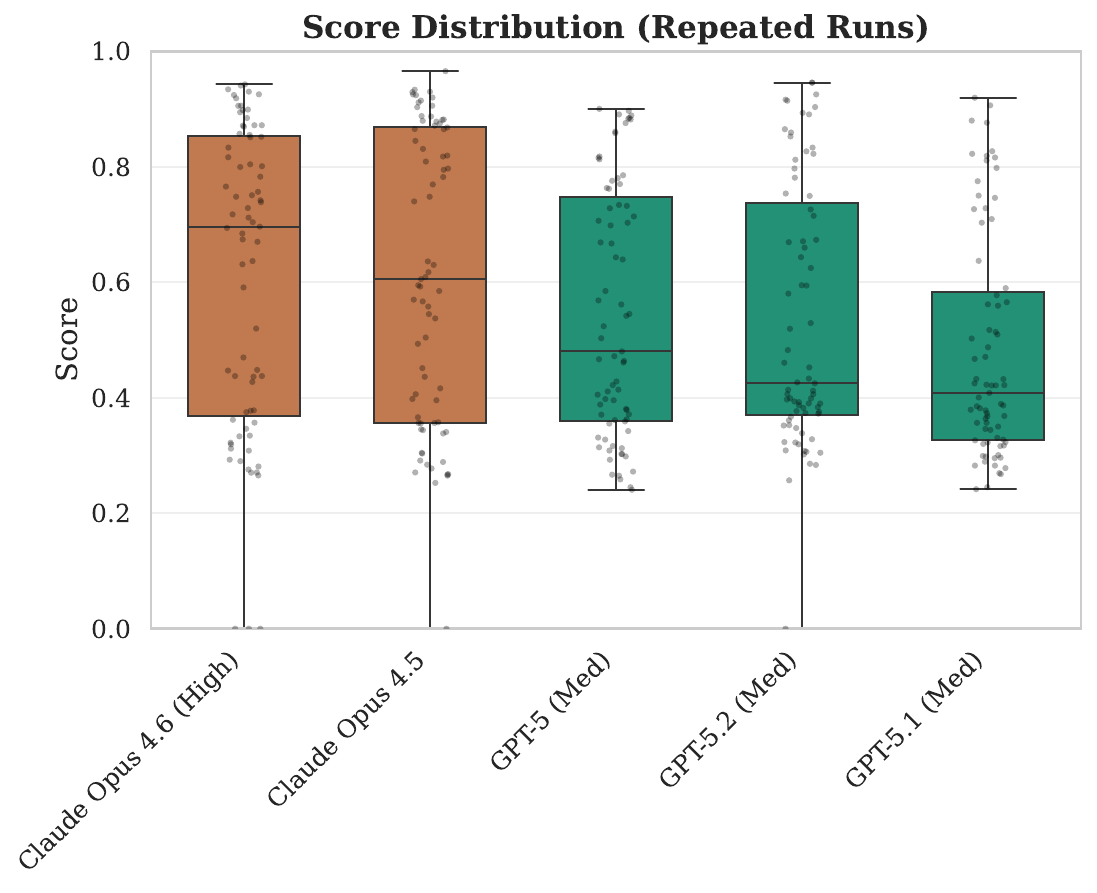}
    \caption{Reward distributions across three repeated runs on CTI-REALM-25.}
    \label{fig:variance}
\end{figure}

\section{Agent Tools}
\label{app:tools}

Table~\ref{tab:agent_tools} describes the eight specialized tools available to agents during evaluation.

\begin{table}[h]
\centering
\caption{Agent tool API.}
\label{tab:agent_tools}
\small
\begin{tabular}{lp{8cm}}
\toprule
Tool & Description \\
\midrule
\texttt{list\_cti\_report\_tags} & Lists all available CTI report tags with metadata \\
\texttt{get\_cti\_reports\_by\_tag} & Retrieves full CTI reports filtered by tag identifier \\
\texttt{list\_kusto\_tables} & Lists available telemetry tables and their schemas \\
\texttt{get\_table\_schema} & Returns column definitions for a specific table \\
\texttt{execute\_kql\_query} & Executes KQL queries against the Kusto cluster \\
\texttt{get\_mitre\_techniques} & Retrieves MITRE ATT\&CK technique details by ID \\
\texttt{search\_sigma\_rules} & Searches the Sigma rule database by keyword \\
\texttt{submit\_detection} & Submits final Sigma rule and KQL query for evaluation \\
\bottomrule
\end{tabular}
\end{table}

\section{Evaluation Details}
\label{app:eval_details}

\textbf{Ground Truth Structure.} Each task includes a structured ground truth object with three components:

\begin{itemize}\setlength{\itemsep}{1pt}
    \item \textbf{\texttt{mitre\_techniques}} --- A list of expected MITRE ATT\&CK technique IDs (e.g., \texttt{["T1115"]}). Used for C1 scoring via Jaccard similarity between the agent's identified techniques and the ground truth set.
    \item \textbf{\texttt{data\_sources}} --- A list of expected telemetry tables the agent should explore (e.g., \texttt{["deviceprocessevents"]}). Used for C2 scoring via Jaccard similarity between tables the agent queried or inspected and the expected set.
    \item \textbf{\texttt{regex\_patterns}} --- A dictionary mapping field names to regex patterns that capture key indicators of compromise (e.g., \texttt{\{"filename": "wl-paste|xclip|xsel"\}}). Used for C4 F1 scoring: each row in the agent's KQL query results is matched against these patterns. A row is a true positive if all field patterns match; precision, recall, and F1 are computed across all returned rows.
\end{itemize}

\noindent The complete structure per task is:
\begin{verbatim}
{"ground_truth": {
    "mitre_techniques": ["T1115"],
    "data_sources": ["deviceprocessevents"],
    "regex_patterns": {
      "filename": "wl-paste|xclip|xsel",
      "initiatingprocessfilename": "bash|sh|dash"
    }
}}
\end{verbatim}

\textbf{LLM-as-Judge Configuration.} C0 (CTI report analysis) and C4 (Sigma rule quality) use GPT-5-Mini as an automated judge. For C0, the judge evaluates tag relevance, report selection appropriateness, and reasoning quality on a 0--1 scale. For Sigma quality, the judge assesses syntax correctness (weight 0.25) and detection specificity (weight 0.75), producing a weighted quality reward.

\section{Agent Prompts}
\label{app:prompts}

\subsection{Standard Agent Prompt}

The following system prompt is used for all standard evaluations (CTI-REALM-25, CTI-REALM-50) and the ablation study. In the ablation variant, the CTI-specific tools (\texttt{list\_cti\_report\_tags}, \texttt{get\_cti\_reports\_by\_tag}) are removed from the agent's toolkit, but the prompt text remains identical.

\begin{tcolorbox}[colback=gray!5, colframe=gray!50, title=Standard Agent System Prompt, fonttitle=\small\bfseries, breakable]
\small\ttfamily
You are a cybersecurity analyst creating detection rules from threat intelligence.

\vspace{4pt}
\#\# Mission\\
1. **Find and analyze relevant CTI report**: Search available threat intelligence reports and explain why the selected report relates to your detection objective\\
2. **Create detection capability** with: (1) Sigma rule, (2) KQL query, (3) Query results

\vspace{4pt}
\#\# Resources\\
You have access to various tools for threat intelligence analysis, detection rule development, and data querying. Use your available tools to explore capabilities and complete the mission.

\vspace{4pt}
\#\# Output Format (JSON only, no extra text)\\
\{\\
\quad"sigma\_rule": "<YAML string with \textbackslash n for line breaks>",\\
\quad"kql\_query": "<working KQL query string>",\\
\quad"query\_results": [\{"column1": "value1", ...\}, ...]\\
\}

\vspace{4pt}
Note: query\_results must be the actual array returned by execute\_kql\_query.
\end{tcolorbox}

\subsection{Memory-Augmented Agent Prompt}

The memory variant adds a \texttt{Memory} section to the standard prompt, instructing the agent to consult pre-loaded guidance files before beginning its investigation.

\begin{tcolorbox}[colback=blue!3, colframe=blue!40, title=Memory Augmentation Addition (inserted between Resources and Output Format), fonttitle=\small\bfseries]
\small\ttfamily
\#\# Memory\\
You have a `memory` tool with pre-loaded guidance in /memories/. **Before starting any investigation, use the memory tool to view /memories/ and read the files there.**

\vspace{4pt}
They contain workflow steps, tool tips, and common patterns that will help you work more effectively. Refer back to your memory notes when you get stuck.
\end{tcolorbox}

\section{LLM-as-Judge Prompts}
\label{app:judge}

\subsection{C0: CTI Threat Alignment Judge}

This judge evaluates whether the agent's threat intelligence research is relevant to the detection objective and whether the agent clearly explained the connection. The judge scores on a 0.0--1.0 scale.

\begin{tcolorbox}[colback=gray!5, colframe=gray!50, title=C0 Judge System Prompt, fonttitle=\small\bfseries]
\small\ttfamily
You are evaluating whether an AI agent's threat intelligence research is relevant to the detection objective and whether they clearly explained the connection between selected CTI reports and the detection task.
\end{tcolorbox}

\begin{tcolorbox}[colback=gray!5, colframe=gray!50, title=C0 Judge Instructions (abbreviated), fonttitle=\small\bfseries, breakable]
\small
Consider these aspects:
\begin{enumerate}
    \item Did the agent search for tags that match the detection objective?
    \item Did they identify specific CTI reports or threat intelligence?
    \item Did they explain WHY the reports are relevant to the detection task?
\end{enumerate}

\vspace{4pt}
\textbf{Scoring calibration (from few-shot examples):}
\begin{itemize}\setlength{\itemsep}{1pt}
    \item \textbf{1.0}: Highly relevant tags, specific report found, outstanding reasoning connecting report to objective with MITRE technique references
    \item \textbf{0.8}: Relevant tags, good report selection, solid reasoning about relevance
    \item \textbf{0.5}: Adequate tags, reports found but not specified, very weak reasoning
    \item \textbf{0.2}: Mismatched tags, no evidence of report analysis, no explanation of relevance
\end{itemize}
\end{tcolorbox}

\subsection{C4: Sigma Rule Quality Judge}

This judge evaluates Sigma rules for production readiness across two dimensions: syntax correctness (weight 0.25) and detection specificity (weight 0.75).

\begin{tcolorbox}[colback=gray!5, colframe=gray!50, title=C4 Sigma Judge System Prompt, fonttitle=\small\bfseries]
\small\ttfamily
You are a Sigma rule evaluator.
\end{tcolorbox}

\begin{tcolorbox}[colback=gray!5, colframe=gray!50, title=C4 Sigma Judge Instructions (abbreviated), fonttitle=\small\bfseries, breakable]
\small
Evaluate Sigma rules for production readiness:

\vspace{4pt}
\textbf{Syntax score} (0.0--1.0, weight 0.25):
\begin{itemize}\setlength{\itemsep}{1pt}
    \item Is the YAML valid and parseable?
    \item Are critical fields present (title, logsource, detection, condition)?
    \item Are recommended fields included (id, status, description, level)?
\end{itemize}

\vspace{4pt}
\textbf{Specificity score} (0.0--1.0, weight 0.75):
\begin{itemize}\setlength{\itemsep}{1pt}
    \item Does the rule match the detection objective precisely?
    \item Are there specific conditions that reduce false positives?
    \item Would this rule be effective in a production environment?
\end{itemize}

\vspace{4pt}
\textbf{Scoring calibration (from few-shot examples):}
\begin{itemize}\setlength{\itemsep}{1pt}
    \item \textbf{Syntax 1.0 / Specificity 0.95}: All required fields, proper YAML, MITRE tags, specific operation names for target behavior
    \item \textbf{Syntax 1.0 / Specificity 0.75}: All required fields, targets correct operation but could add contextual filters
    \item \textbf{Syntax 0.7 / Specificity 0.2}: Missing metadata fields; triggers on broad category (e.g., any PowerShell) rather than specific behavior
    \item \textbf{Syntax 0.3 / Specificity 0.0}: Missing detection section entirely
    \item \textbf{Syntax 0.1 / Specificity 0.0}: Malformed YAML, cannot parse
\end{itemize}

\vspace{4pt}
Final Sigma quality score: $0.25 \times \text{syntax} + 0.75 \times \text{specificity}$
\end{tcolorbox}


\section{Token Usage}
\label{app:tokens}

\begin{table}[h]
\centering
\caption{Per-sample token usage breakdown on CTI-REALM-50, sorted by reward.}
\label{tab:token_usage}
\small
\begin{tabular}{lrrrr}
\toprule
Model & Input & Output & Reasoning & Total \\
\midrule
Claude Opus 4.6 & 2,269 & 7,116 & --- & 472,352 \\
Claude Opus 4.5 & 2,437 & 7,434 & --- & 441,915 \\
Claude Sonnet 4.5 & 71 & 9,528 & --- & 523,693 \\
GPT-5 (Med) & 326,535 & 9,975 & 4,858 & 336,510 \\
GPT-5.2 (Med) & 367,329 & 8,129 & 3,095 & 375,457 \\
GPT-5 (High) & 390,512 & 11,990 & 6,729 & 402,501 \\
GPT-5.2 (High) & 528,312 & 11,022 & 5,701 & 539,333 \\
GPT-5 (Low) & 172,250 & 6,863 & 1,857 & 179,113 \\
GPT-5.1 (Med) & 243,210 & 14,054 & 7,745 & 257,265 \\
GPT-5.1 (High) & 420,308 & 24,100 & 17,359 & 444,408 \\
GPT-5.2 (Low) & 220,518 & 5,432 & 1,123 & 225,950 \\
O3 & 360,962 & 7,685 & 3,259 & 368,647 \\
GPT-5-Mini & 238,206 & 14,056 & 5,958 & 252,263 \\
GPT-4.1 & 116,855 & 3,468 & --- & 120,323 \\
GPT-5.1 (Low) & 135,373 & 6,077 & 1,026 & 141,450 \\
O4-Mini & 185,854 & 6,931 & 4,214 & 192,784 \\
\bottomrule
\end{tabular}
\end{table}

\section{Example Agent Trajectory}
\label{app:trajectory}

The following shows an abbreviated trajectory from Claude Opus 4.6 (High) on a Linux endpoint task (sample \texttt{linux\_041}, reward: 0.552). The agent completed the task in 11 reasoning steps across 36 messages.

\begin{tcolorbox}[colback=gray!3, colframe=gray!50, title={\textbf{Task Objective}}, fonttitle=\small\bfseries, top=2pt, bottom=2pt]
\scriptsize
Write a detection rule to identify scheduled task creation using \texttt{at} command by monitoring for shell commands piped to \texttt{at}, indicating attempts to establish persistence or delayed execution of malicious commands.
\end{tcolorbox}

\vspace{-4pt}

\begin{tcolorbox}[colback=blue!2, colframe=blue!30, fonttitle=\small\bfseries, title=Agent Steps, breakable, top=2pt, bottom=2pt]
\scriptsize

\textbf{Step 1} --- \texttt{list\_cti\_report\_tags}, \texttt{search\_mitre\_techniques}, \texttt{list\_kusto\_tables}\\
\textit{``I'll start by researching the threat landscape, available data, and relevant detection techniques for scheduled task creation using the at command.''}\\
$\rightarrow$ Found T1053 (Scheduled Task/Job) as the relevant MITRE technique; listed available tables.

\medskip
\textbf{Step 2} --- \texttt{get\_cti\_reports\_by\_tag} $\times$2, \texttt{search\_sigma\_rules} $\times$2, \texttt{get\_table\_schema}\\
\textit{``T1053 is the relevant technique. Let me get CTI reports for scheduled tasks and examine the data tables.''}\\
$\rightarrow$ Retrieved CTI reports on Linux crontab; no Sigma rules for \texttt{at} specifically; obtained \texttt{DeviceProcessEvents} schema.

\medskip
\textbf{Step 3} --- \texttt{get\_cti\_reports\_by\_tag}, \texttt{search\_sigma\_rules} $\times$2, \texttt{sample\_table\_data}\\
\textit{``Let me search for more specific CTI reports and Sigma rules, and look at the data.''}\\
$\rightarrow$ Found relevant CTI report from Elastic Security Labs covering T1053.002; sampled telemetry.

\medskip
\textbf{Step 4} --- \texttt{search\_sigma\_rules} $\times$2, \texttt{execute\_kql\_query}\\
\textit{``Found the Elastic report specifically covering T1053.002 (At). Let me search for existing Sigma rules.''}\\
$\rightarrow$ Found existing Sigma rule ``Scheduled Task/Job At''; initial query returned \texttt{at} command activity.

\medskip
\textbf{Step 5} --- \texttt{execute\_kql\_query}\\
\textit{``Found data with at command usage. Let me run a more focused query for piped commands to at.''}\\
$\rightarrow$ Identified specific patterns in process command lines.

\medskip
\textbf{Step 6--9} --- \texttt{execute\_kql\_query} $\times$4\\
\textit{``Iteratively refining queries to capture pipe-to-at patterns, file input (-f flag), and env wrapper patterns.''}\\
$\rightarrow$ Found key attack pattern: \texttt{echo 'command' | at now + 3 minutes}; identified three distinct attack variants.

\medskip
\textbf{Step 10} --- \texttt{execute\_kql\_query}\\
\textit{``Building the comprehensive detection query covering all three attack patterns.''}\\
$\rightarrow$ Final query returned 10 results covering pipe-to-at, file input, and env wrapper patterns.

\medskip
\textbf{Step 11} --- \texttt{validate\_output\_json}\\
\textit{``Validating the final output JSON.''}\\
$\rightarrow$ Output validated successfully.

\end{tcolorbox}

\vspace{-4pt}

\begin{tcolorbox}[colback=green!3, colframe=green!40!black, fonttitle=\small\bfseries, title=Final Output (abbreviated), top=2pt, bottom=2pt]
\scriptsize
\textbf{Sigma Rule:}
\begin{verbatim}
title: Scheduled Task Creation via Shell Command Piped to At
status: experimental
description: Detects scheduled task creation using 'at'
  where shell commands are piped to 'at' via echo/printf/cat,
  or 'at' is invoked with -f flag for script execution.
tags: [attack.persistence, attack.t1053.002]
logsource: {category: process_creation, product: linux}
detection:
  selection:
    Image|endswith: '/at'
    CommandLine|contains|any: ['| at ','at -f','-c at']
  condition: selection
level: medium
\end{verbatim}
\vspace{-4pt}
\textbf{KQL Query:}
\begin{verbatim}
DeviceProcessEvents
| where (ProcessCommandLine matches regex
    @"\|\s*at\s+(now|noon|midnight|\d{1,2}:\d{2})")
  or (ProcessCommandLine has "| at "
    and ProcessCommandLine has_any
    ("echo","printf","cat","bash","sh"))
  or (FolderPath endswith "/at")
| project Timestamp, DeviceName, AccountName,
    FileName, ProcessCommandLine,
    InitiatingProcessFileName
\end{verbatim}
\end{tcolorbox}

\vspace{-4pt}

\begin{tcolorbox}[colback=purple!3, colframe=purple!40, fonttitle=\small\bfseries, title=Scoring Breakdown, top=2pt, bottom=2pt]
\scriptsize
Rewards shown as weighted checkpoint rewards ($w_i \times r_i$ scaled to 10 points); individual $r_i \in [0,1]$.\\
\begin{tabular}{llll}
\textbf{C0} CTI Report Analysis: 1.12/1.25 &
\textbf{C1} MITRE Mapping: 0.75/0.75 &
\textbf{C2} Data Exploration: 1.00/1.00 &
\textbf{C3} Query Execution: 0.50/0.50 \\
\end{tabular}\\[2pt]
\textbf{C4} Detection Quality: 2.65/6.50 (F1: 0.35, Sigma: 0.61) \quad
\textbf{Total:} 5.52/10.0 $\rightarrow$ \textbf{Normalized Reward: 0.552}
\end{tcolorbox}

\section{Seed Memory Content}
\label{app:seed_memory}

The memory augmentation study (Section~\ref{sec:memory}) provides the agent with three pre-loaded guidance files. Excerpts from each are shown below.

\subsection{Workflow Guidance}
\label{app:seed_workflow}

Provides a structured step-by-step detection engineering workflow.

\begin{tcolorbox}[colback=blue!3, colframe=blue!40, title=workflow.md (excerpt), fonttitle=\small\bfseries, breakable]
\small\ttfamily
\# CTI Detection Rule Development Workflow\\[4pt]
\#\# Critical: Always Start With CTI Reports\\
Successful investigations begin by understanding the threat\\
BEFORE exploring data. Do NOT jump straight to data tables.\\[4pt]
\#\# Recommended Steps\\
1. \textbf{Discover CTI reports} --- use list\_cti\_report\_tags()\\
\quad to see available tags, then get\_cti\_reports\_by\_tag(tag)\\
\quad to pull reports matching your detection objective.\\
2. \textbf{Map MITRE techniques} --- use\\
\quad search\_mitre\_techniques(tactic) to find technique IDs.\\
3. \textbf{Search Sigma rules} --- use search\_sigma\_rules()\\
\quad to find existing detection logic to adapt.\\
4. \textbf{Explore data} --- use list\_kusto\_tables(),\\
\quad get\_table\_schema(table), sample\_table\_data(table).\\
5. \textbf{Write \& test KQL} --- use execute\_kql\_query(query).\\
6. \textbf{Build Sigma rule} --- write a YAML Sigma rule.\\
7. \textbf{Validate \& submit} --- call validate\_output\_json().
\end{tcolorbox}

\subsection{Tool Usage Tips}
\label{app:seed_tools}

Provides specific guidance on each tool function, including common failure modes and debugging strategies.

\begin{tcolorbox}[colback=blue!3, colframe=blue!40, title=tool\_tips.md (excerpt), fonttitle=\small\bfseries, breakable]
\small\ttfamily
\# Tool Usage Tips\\[4pt]
\#\# CTI Report Tools (use these FIRST)\\
- list\_cti\_report\_tags(): Returns all available tags with\\
\quad report counts. Start here to find matching tags.\\
- get\_cti\_reports\_by\_tag(tag): Returns full reports with\\
\quad id, title, link, tags, and content.\\[4pt]
\#\# Kusto/KQL Tools\\
- list\_kusto\_tables(): Discover available tables.\\
\quad Do not guess table names.\\
- get\_table\_schema(table): Get exact column names\\
\quad and types. Always check BEFORE writing any query.\\
- execute\_kql\_query(query): If 0 rows returned:\\
\quad 1. Check column names match the schema exactly\\
\quad 2. Use `contains` instead of `==` for string filters\\
\quad 3. Remove time filters to broaden results
\end{tcolorbox}

\subsection{Common Patterns}
\label{app:seed_patterns}

Provides template patterns for Sigma rules and KQL queries, plus an investigation decision tree for debugging.

\begin{tcolorbox}[colback=blue!3, colframe=blue!40, title=patterns.md (excerpt), fonttitle=\small\bfseries, breakable]
\small\ttfamily
\# Common Patterns\\[4pt]
\#\# Sigma Rule Skeleton\\
title: <Detection Name>\\
status: experimental\\
tags:\\
\quad - attack.<tactic>\\
\quad - attack.<technique\_id>\\
logsource:\\
\quad category: <process\_creation | file\_event | ...>\\
\quad product: <windows | linux | azure>\\
detection:\\
\quad selection:\\
\quad\quad FieldName|<modifier>: <value>\\
\quad condition: selection\\[4pt]
\#\# Investigation Decision Tree\\
- Query returns 0 rows?\\
\quad $\rightarrow$ Check column names against schema\\
\quad $\rightarrow$ Broaden filters: contains instead of ==\\
\quad $\rightarrow$ Sample the table to see actual values\\
- Query returns errors?\\
\quad $\rightarrow$ Verify table name via list\_kusto\_tables()\\
\quad $\rightarrow$ Check KQL syntax
\end{tcolorbox}

\end{document}